\documentclass[12pt]{article}
\usepackage{epsfig}
\usepackage{amsfonts}
\usepackage{latexsym}
\usepackage{amsmath}
\usepackage{mathrsfs}
\usepackage{hyperref}
\numberwithin{equation}{section}

\DeclareFontFamily{OT1}{rsfs}{}
\DeclareFontShape{OT1}{rsfs}{m}{n}{
<-7> rsfs5 <7-10> rsfs7 <10-> rsfs10}{}
\DeclareMathAlphabet{\mycal}{OT1}{rsfs}{m}{n}

\newcommand{\bea}{\begin{eqnarray}}
\newcommand{\eea}{\end{eqnarray}}
\def\non{\nonumber}
\def\d{\partial}

\begin{document}

\begin{titlepage}
\unitlength = 1mm
\begin{center}

{ \LARGE {\textsc{Fast plunges into Kerr black holes}}}

\vspace{1.8cm}
Shahar Hadar$^\S$, Achilleas P. Porfyriadis$^\dag$ and Andrew Strominger$^\dag$

\vspace{1cm}

{\it $^\S$ Racah Institute of Physics, Hebrew University, Jerusalem 91904, Israel}

\vspace{0.5cm}

{\it $^\dag$ Center for the Fundamental Laws of Nature, Harvard University,\\
Cambridge, MA 02138, USA}

\vskip 2cm

\begin{abstract}
Most extreme-mass-ratio-inspirals of small compact objects into supermassive black holes end with a fast plunge from an eccentric last stable orbit. For rapidly rotating black holes such fast plunges may be studied in the context of the Kerr/CFT correspondence because they occur in the near-horizon region where dynamics are governed by the infinite dimensional conformal symmetry. In this paper we use conformal transformations to analytically solve for the radiation emitted from fast plunges into near-extreme Kerr black holes. We find perfect agreement between the gravity and CFT computations.
\end{abstract}

\vspace{1.0cm}

\end{center}
\end{titlepage}

\pagestyle{plain}
\setcounter{page}{1}
\newcounter{bean}
\baselineskip18pt


\setcounter{tocdepth}{2}

\tableofcontents

\section{Introduction}

Properties of diffeomorphisms in general relativity imply that gravitational dynamics near the horizons of rapidly-rotating Kerr black holes (BHs) are constrained by an infinite-dimensional conformal symmetry \cite{hep-th/9905099,0809.4266}.  We refer to this as the `weak' Kerr/CFT correspondence. It enables powerful analytic techniques developed in the study of two-dimensional conformal field theory (CFT) to be employed in the analysis of the near-horizon gravitational dynamics.  The conjectured `strong' Kerr/CFT correspondence (see \cite{Bredberg:2011hp,Compere:2012jk} for reviews) is that quantum gravity in the near-horizon region of such BHs is dual to a (warped) two-dimensional conformal field theory. This conjecture is relevant for the study of the quantum puzzles surrounding black holes.
However, for some interesting questions strong Kerr/CFT is not needed.  In particular weak Kerr/CFT is sufficient for questions which arise in observational astronomy \cite{McClintock:2006xd, Brenneman:2006hw, Gou:2013dna}.

One such question refers to gravitational radiation production by extreme-mass-ratio-inspirals (EMRIs) into near-extremal Kerr BHs. These systems are binaries which consist of a small compact object inspiralling into a much larger BH and are important candidates for direct gravitational wave (GW) detection by eLISA \cite{Finn:2000sy,Gair:2004iv,elisa}.  They are the focus of this paper, the third in a series  \cite{Porfyriadis:2014fja,Hadar:2014dpa}. For certain initial conditions, such as those furnished by separation events of binary systems passing near a supermassive BH, EMRIs consist of two stages \cite{AmaroSeoane:2007aw, Hadar:2009ip}: (i) a long quasi-circular inspiral followed by (ii) a quick plunge into the black hole from the innermost stable circular orbit (ISCO) at the edge of the accretion disk. For nearly extremal black holes, both near-ISCO circular orbits and the post-ISCO plunge take place deep in the near horizon region where Kerr/CFT predicts the extra conformal symmetry. More precisely, the latest part of stage (i) takes place in the near-horizon extreme Kerr (NHEK) geometry while the entire stage (ii) takes place in the near-horizon near-extreme Kerr (near-NHEK) geometry. Global $SL(2,R)$ conformal symmetry and local Virasoro conformal symmetry have been utilized recently in \cite{Porfyriadis:2014fja} and \cite{Hadar:2014dpa} to calculate the corresponding emitted gravitational waves during a near-ISCO circular orbit and the post-ISCO plunge, respectively. The analysis was  entirely analytic -- see \cite{Taracchini:2014zpa,Nagar:2014kha} for recent and compatible numerical analyses of this problem.\footnote{The $SL(2,R)$ isometries have also been exploited in recent studies of BH magnetospheres \cite{Lupsasca:2014pfa,Zhang:2014pla,Lupsasca:2014hua} in order to find new analytical solutions to the equations of force-free electrodynamics in the extreme Kerr background.}

Despite the fact that radiation reaction circularizes trajectories rather efficiently, most astrophysical relevant EMRIs are expected to have such large initial eccentricities that more than 50\% of all observable EMRIs will end with plunges from a moderately eccentric last stable orbit \cite{Barack:2003fp}. Thus in this paper we extend the analysis of \cite{Hadar:2014dpa}, where only the special post-ISCO plunge was analyzed, and compute analytically the radiative signature of a new family of trajectories we call ``fast plunges''. These plunges are trajectories with arbitrary (near-)NHEK orbital energy which start at some finite time from the (near-)NHEK boundary. This is in contrast with the special post-ISCO plunge of \cite{Hadar:2014dpa} which comes in from the near-NHEK boundary in the infinite past. The fact that the fast plunges are soluble is again a direct consequence of the conformal symmetry: we show that they are related by conformal mapping to the circular orbit of \cite{Porfyriadis:2014fja}. In this paper we study a simplified version of the problem, replacing gravitational perturbations with a scalar field. The analysis is, however, readily generalizable to the linearized gravity case.

The holographic dual of the gravity problem with the geodesic particle on a circular orbit sourcing a bulk field may be understood as follows. The dual field theory is a CFT which is subject to an external driving source for the operator dual to the bulk field. The source is driving the system at a frequency given by the orbital frequency of the geodesic particle and induces a constant transition rate out of the initial state which matches exactly the bulk particle number flux down the BH horizon \cite{Porfyriadis:2014fja}. The holographic dual of the plunge problem has a richer dynamical evolution involving a quantum quench of the CFT that injects energy into the system which is subsequently thermalized and equilibrated.  On the gravity side, the approach to equilibrium is characterized at late times by the quasinormal modes. As in \cite{Hadar:2014dpa}, in this paper we will only consider holographically the very high frequency limit, that is to say, we will only consider timescales which are much shorter than that of the quench. In this case, we can again think of the CFT as being driven continuously by an external source throughout the entire period under study. The corresponding transition rates may then be obtained by conformal transformations in the CFT and we will see that the results remain in perfect agreement with the gravity calculations.

This paper is organized as follows. In section \ref{Fast NHEK plunge} we analyze the fast NHEK plunge. In \ref{trajectory and mapping} we present the plunge trajectory and the mapping which takes it to a circular orbit. In \ref{solution in NHEK} we use this mapping to compute the radiation emitted in the plunge. In \ref{CFT analysis} we perform the dual conformal field theory computation, and show the two agree. In \ref{matching to far region} we reattach the asymptotically flat region and determine the radiation at future null infinity. In section \ref{Fast near-NHEK plunge} we repeat the above steps for the fast near-NHEK plunge.

\section{Radiation from the fast NHEK plunge}\label{Fast NHEK plunge}

The Kerr metric in Boyer-Lindquist coordinates is given by
\bea
ds^2 = -\frac{\Delta}{\hat{\rho}^2}\left( d\hat{t} - a \sin^2\theta \, d\hat{\phi} \right)^2 + \frac{\sin^2\theta}{\hat{\rho}^2} \left( (\hat{r}^2 + a^2)d\hat{\phi} - a \, d\hat{t} \right)^2+\frac{\hat{\rho}^2}{\Delta}d\hat{r}^2 + \hat{\rho}^2 d\theta^2 \, \, ,
\label{Kerr}
\eea
where
\bea
\Delta = \hat{r}^2 - 2 M \hat{r} + a^2 \, \, , ~~~~ \hat{\rho}^2 = \hat{r}^2 + a^2 \cos^2 \theta \, \, ,
\label{Kerr defs}
\eea
and is characterized by the BH mass $M$ and angular momentum $J = a M$. The outer/inner horizons are situated at $r_{\pm} = M \pm \sqrt{M^2-a^2}$. For extremal BHs $a = M$ and the NHEK geometry may be obtained by making the coordinate transformation
\bea
r = \frac{\hat{r}-r_+}{\epsilon \, r_+} \, \, , ~~~~ \, \, t = \frac{\epsilon \, \hat{t}}{2 M} \, \, , ~~~~ \, \, \phi = \hat{\phi} - \frac{\hat{t}}{2 M} \, \, ,
\label{NHEK zoom}
\eea
and taking the limit $\epsilon\to 0$, keeping $r,t$ finite. This effectively zooms into the near horizon region and yields the non-singular NHEK metric \cite{hep-th/9905099}:
\bea
ds^2 = 2M^2 \Gamma(\theta)\left[ -r^2 dt^2 + \frac{dr^2}{r^2} + d\theta^2 + \Lambda(\theta)^2 (d\phi + r dt)^2 \right] \, \, ,
\label{NHEK 1}
\eea
where
\bea
\Gamma(\theta) = \frac{1+\cos^2\theta}{2} \, \, , ~~~~ \Lambda(\theta) = \frac{2\sin\theta}{1+\cos^2\theta} \, \, .
\label{lambda and gamma}
\eea

\subsection{Trajectory \& mapping}\label{trajectory and mapping}
Consider the equatorial plunging trajectories of small (test) compact objects in NHEK with energy and angular momentum (per unit rest mass) given by
\bea
e=\frac{4M}{\sqrt{3} R_0} \, \, , ~~~~ l = \frac{2M}{\sqrt{3}} \, \, ,
\label{e and l}
\eea
for arbitrary $R_0 > 0$.
The solution for this trajectory is a fast NHEK plunge:
\bea
t(r) &=& \frac{1}{r}\sqrt{1+R_0 r} \, \, ,  \non \\
\phi(r) &=& \frac{3}{4} \sqrt{1+R_0 r} + \ln\frac{\sqrt{1+R_0 r}-1}{\sqrt{1+R_0 r}+1} + \Phi_0 \, \, .
\label{trajectory solution}
\eea
\begin{figure}[!hb]
\includegraphics[angle=0, width=1.6in]{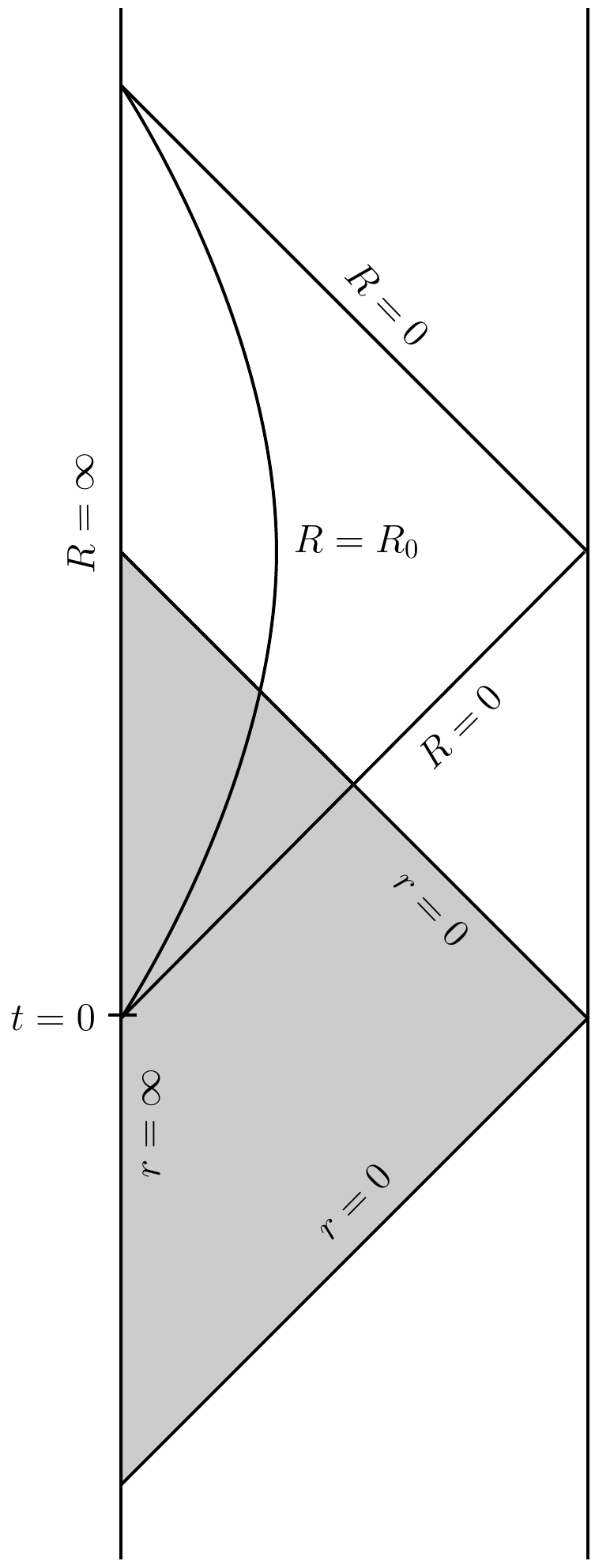}\centering
\caption{Penrose diagram with the fast NHEK plunge. Both upper (bounded by $R=0$ and $R=\infty$) and lower (bounded by $r=0$ and $r=\infty$, shaded) wedges are NHEK. The line $R=R_0$ is a circular orbit in the upper wedge. In the shaded wedge it plunges from the boundary $r=\infty$ at $t=0$ into the future horizon $r=0$.}\label{NHEK diagram}
\end{figure}
The mapping
\bea
T &=& - \frac{r^2 t}{r^2 t^2 - 1} \, \, ,   \non \\
R &=& \frac{r^2 t^2-1}{r} \, \, ,  \label{mapping fast NHEK}  \\
\Phi &=& \phi + \ln\frac{rt+1}{rt-1}  \non \, \, ,
\eea
maps the $rt \geq 1$ part of the NHEK patch (\ref{NHEK 1}) to the $RT \leq -1$ part of another NHEK patch (see Figure \ref{NHEK diagram}),
\bea
ds^2 = 2M^2 \Gamma(\theta)\left[ -R^2 dT^2 + \frac{dR^2}{R^2} + d\theta^2 + \Lambda(\theta)^2 (d\Phi + R dT)^2 \right] \, \, .
\label{NHEK 2}
\eea
The fast plunge trajectory (\ref{trajectory solution}) is mapped under \eqref{mapping fast NHEK} to
\bea
R &=& R_0 \, \, , \non \\
\Phi &=& -\frac{3}{4} R_0 T + \Phi_0 \, \, .
\label{circular orbit}
\eea
The above, being a circular orbit, is a much simpler problem which has been solved, both in gravity and the CFT, in \cite{Porfyriadis:2014fja}.
Below we will use the transformation \eqref{mapping fast NHEK} in order to solve for the radiation produced during the fast plunge (\ref{trajectory solution}).

\subsection{Gravity analysis}\label{solution in NHEK}

Consider a star carrying scalar charge. The action for this system is given by
\bea
S = -\frac{1}{2}\int d^4x \sqrt{-g} \left[ \left( \partial \Psi \right)^2 + 8 \pi \lambda  \Psi \mathcal{S} \right] \, \, ,
\label{scalar action}
\eea
where $\lambda$ is a coupling constant and
\bea
\mathcal{S}(x) = -\int d\tau (-g)^{-1/2} \delta^{(4)}(x-x_* (\tau)) \, \, ,
\label{scalar source}
\eea
is the source term, with $x_* (\tau)$ the trajectory (\ref{trajectory solution}).

The solution for a star in a circular orbit in NHEK (\ref{circular orbit}), with ingoing boundary condition at the horizon and Neumann at infinity, is (setting $\Phi_0=0$) \cite{Porfyriadis:2014fja}:
\bea
\Psi = \sum_{\ell, m} e^{i m (\Phi + 3 R_0 T/4)} S_\ell(\theta)R^{(c)}_{\ell m}(R) \, \, ,
\label{circular orbit solution}
\eea
where $S_\ell$ are NHEK spheroidal harmonics obeying
\bea
\frac{1}{\sin \theta} \d_{\theta} (\sin \theta \d_{\theta} S_{\ell}) + \left( K_{\ell} - \frac{m^2}{\sin^2 \theta} - \frac{m^2}{4} \sin^2 \theta \right)S_{\ell} = 0 \, \, ,
\label{extremal spheroidal harmonic equation}
\eea
and
\bea
R^{(c)}_{\ell m} = \frac{1}{W_{\Omega}} \left[ X \, \Theta(R_0-R) W_{im,h-1/2}\left( {-2i\Omega}/{R} \right) + Z \, \Theta(R-R_0) M_{im,h-1/2}\left( {-2i\Omega}/{R} \right) \right] .
\label{radial wavefunction circular}
\eea
In the above $W_{k,\mu}$, $M_{k,\mu}$ are Whittaker functions and
\bea
\Omega &=& -\frac{3 m R_0}{4} \, \, , \non \\
h &=& \frac{1}{2} + \sqrt{\frac{1}{4} + K_\ell - 2m^2} \, \, , \non \\
X &=& -\frac{\sqrt{3} \lambda R_0}{2 M} S_{\ell}\left( {\pi}/{2} \right) M_{im,h-1/2}\left( {3im}/{2} \right) \, \, , \non \\
Z &=& -\frac{\sqrt{3} \lambda R_0}{2 M} S_{\ell}\left( {\pi}/{2} \right) W_{im,h-1/2}\left( {3im}/{2} \right) \, \, ,  \non \\
W_{\Omega} &=& 2i\Omega \frac{\Gamma(2h)}{\Gamma(h-im)} \, \, .
\label{circular solution defs}
\eea

We will now compute the radiation produced by a star plunging into the black hole on the trajectory (\ref{trajectory solution}). As shown in section \ref{trajectory and mapping}, this orbit is related to a circular orbit by \eqref{mapping fast NHEK}. We use this mapping to generate the solution for fast plunge radiation directly from (\ref{circular orbit solution}). It is important to note that the boundary condition of no incoming radiation from the past horizon in (\ref{NHEK 1}) implies no incoming radiation from the past horizon in (\ref{NHEK 2}), since for $rt<1$ in (\ref{NHEK 1}) $\Psi=0$ (see Figure \ref{NHEK diagram}).

\noindent
\textbf{Solution at the boundary.}
For fixed $t,\phi$ and $r \to \infty$
\bea
R &\approx& r \, t^2 \to \infty \, \, , \non \\
T &\approx& -\frac{1}{t} \, \, , \non \\
\Phi &\approx& \phi \, \, .
\label{mapping infinity}
\eea
Using
\bea
\left. M_{im,h-1/2}\left( {-2i\Omega}/{R} \right)\right|_{R \to \infty} \, \, \to \, \, \, \, (-2i\Omega)^h R^{-h} \, \, ,
\label{whittaker M infty}
\eea
we obtain
\bea
\left. R^{(c)}_{\ell m} \right|_{r \to \infty} &\approx& \frac{Z}{W_{\Omega}} (-2i\Omega)^h t^{-2h} r^{-h} \, \, ,
\label{R and exp at boundary}
\eea
Putting this together, the solution (\ref{circular orbit solution}) near the boundary is
\bea
\left. \Psi \right |_{r \to \infty} = \Theta(t) \sum_{\ell, m} e^{im\left( \phi-\frac{3R_0}{4t} \right)} S_{\ell}(\theta)\left[ \frac{Z}{W_{\Omega}} (-2i\Omega)^h \right] t^{-2h} r^{-h} \, \, ,
\label{particular sol at boundary}
\eea
where the Heaviside $\Theta$-function is added since $\Psi(t<0) = 0$.
In terms of the Fourier decomposition
\bea
\left. \Psi \right |_{r \to \infty} = \frac{1}{\sqrt{2\pi}} \int d \omega \sum_{\ell, m} e^{i (m \phi - \omega t) } S_{\ell}(\theta)  \, R^\infty_{\ell m\omega}(r) \, ,
\label{particular sol at boundary freq}
\eea
we find
\bea
R^\infty_{\ell m\omega} \, = \,  \sqrt{\frac{2}{\pi}} \left[ \frac{Z}{W_{\Omega}} (-2i\Omega)^h \right] \left( \frac{3mR_0}{4 \omega} \right)^{1/2-h} i e^{i\pi h} \cos(2 \pi h) \, K_{2h-1}\left(\sqrt{3  m  R_0  \omega} \right) \, r^{-h}  \, \, ,
\label{fourier infty}
\eea
where $K_\nu$ is the modified Bessel function of the second kind and we take $\omega>0$ and real $h<1$ for simplicity.

\noindent
\textbf{Solution at future horizon.}
For $r \to 0$ with fixed $v\equiv t-{1}/{r}$ and $\phi$,
\bea
R &\approx& 2v \, \, , \non \\
T &\approx& -\frac{1}{2v} \, \, , \non \\
\Phi &\approx& \phi + \ln\frac{2}{r v} \, \, .
\label{mapping horizon}
\eea
The solution near the horizon, for $v>0$, is given by
\bea
\Psi_{hor} = \sum_{\ell, m}  e^{im \phi} S_{\ell}(\theta) \left( \frac{2}{r  v} \right)^{i m} e^{-\frac{3 i m R_0}{8 v}} R^{(c)}_{\ell m}(2v) \, \, ,
\label{particular sol at horizon}
\eea
while $\Psi(v<0)=0$.
In terms of the Fourier decomposition
\bea
\Psi_{hor} &=& \frac{1}{\sqrt{2\pi}} \int d\omega \sum_{\ell, m}  e^{i(m \phi - \omega t)} S_{\ell}(\theta) \, R^{hor}_{\ell m \omega}(r) \, , \label{fourier horizon1} \\
R^{hor}_{\ell m \omega} &=&  \left( \frac{2}{r} \right)^{i m} e^{i \frac{\omega}{r}} \frac{X}{W_{\Omega}} \frac{1}{\sqrt{2 \pi}} \int_{0}^{\infty} dv\,  e^{i \omega v} v^{-i m} e^{-\frac{3 i m R_0}{8 v}} W_{im,h-1/2}\left( \frac{3 i m R_0}{4 v} \right)   \, \, ,
\label{fourier horizon2}
\eea
where in (\ref{fourier horizon2}) we consider the high frequency limit
$\omega R_0 \gg 1$, so only the $v<{R_0}/{2}$ part contributes to the Fourier transform. The integral in (\ref{fourier horizon2}) can be directly evaluated, then approximated using $\omega R_0 \gg 1$ to yield, for $\omega,m>0$ and real $h<1$,
\bea
R^{hor}_{\ell m \omega} &=&  \left( \frac{2}{r} \right)^{i m} \, e^{i \frac{\omega}{r}} \, \frac{X}{W_{\Omega}} \frac{1}{\sqrt{2 \pi}}\, i \, e^{\frac{\pi m}{2}} \, \omega ^{-\frac{1}{2}+i m} \, \sqrt{3 m R_0}\, K_{2 h-1}\left(\sqrt{3 m R_0 \omega }\right) \non \\
&\approx& {i\over 2} \, \frac{X}{W_{\Omega}} 2^{i m} \, (3 m R_0 \omega)^{\frac{1}{4}} \, \omega ^{-1+i m} \,e^{\frac{\pi  m}{2}} \, e^{- \sqrt{3 m R_0 \omega }}\, r^{-im} \, e^{i \frac{\omega}{r}}  \, \, .
\label{fourier horizon3}
\eea
The particle number flux down the horizon, as obtained by integrating the Klein-Gordon current $J^{\mu} = \frac{i}{8 \pi} \left( \Psi^{*} \nabla \Psi - \Psi \nabla \Psi^{*} \right)$, is given for $\omega R_0 \gg 1$ by
\bea
\mathcal{F}_{\ell m \omega} = -\int\sqrt{-g} J^r d\theta d\phi \approx \frac{M^2}{8 \pi} \left| \frac{X}{W_\Omega} \right|^2 \sqrt{\frac{3 m R_0}{\omega}} e^{\pi m} e^{-2\sqrt{3 m R_0 \omega}} \, \, .
\label{flux}
\eea

\subsection{CFT analysis}\label{CFT analysis}
Holographically, the introduction of a geodesic particle in the NHEK bulk may be understood as a particular deformation of the boundary CFT via an external driving source
\bea
S = S_{CFT} + \sum_{\ell} \int J_{\ell}(\Phi,T) \mathcal{O}_{\ell}(\Phi,T) d\Phi dT \, \, .
\label{CFT sources}
\eea
Here $\mathcal{O}_{\ell}$ are CFT operators with left and right weights $h$ and for the circular orbit (\ref{circular orbit}) it was shown in \cite{Porfyriadis:2014fja} that the source is given by
\bea
J_{\ell} = \sum_{m} \frac{X}{W_{\Omega}} (-2i \Omega)^{1-h} \frac{\Gamma(2h-1)}{\Gamma(h-im)} \, e^{i m \Phi- i \Omega T} \, \, .
\label{circular J}
\eea
It follows from conformal invariance of (\ref{CFT sources}) that $J_{\ell}$ carry left and right weights $1-h$.
The transformation \eqref{mapping fast NHEK} induces the following $SL(2,R)$ conformal transformation on the boundary
\bea
T &=& -\frac{1}{t} \, \, ,  \non \\
\Phi &=& \phi \, \, .
\label{boundary conformal transformation}
\eea
The transformed sources corresponding to the fast plunging star are therefore given by
\bea
J_{\ell}(\phi,t) = \Theta(t) t^{2h-2} \sum_{m} \frac{X}{W_{\Omega}} (-2i \Omega)^{1-h} \frac{\Gamma(2h-1)}{\Gamma(h-im)} \, e^{i m \phi + i {\Omega}/{t}} \, \, ,
\label{plunge J}
\eea
and the step function $\Theta(t)$ is added because the quench is performed at $t=0$ (see figure \ref{NHEK diagram}). As explained in the introduction, in this paper we will only consider the high frequency limit $\omega R_0 \gg 1$ which allows us to apply Fermi's golden rule and compare the constant transition rate out of the vacuum with the particle number flux \eqref{flux}. In this limit, in terms of the Fourier decomposition, the sources are, for $\omega,m>0$ and real $h<1$,
\bea
J_{\ell}(\phi,t) &=& \frac{1}{\sqrt{2 \pi}} \int d\omega \sum_{m} J_{\ell m \omega} e^{i(m\phi - \omega t)} \, \, , \non \\
J_{\ell m \omega} &=& \frac{X}{W_\Omega} \, \frac{\Gamma (2 h-1)}{\Gamma (h-i m)} \, (3 m R_0 \omega)^{1/4} \, (2\omega)^{-h} \, \, e^{ \frac{i \pi h}{2} - \sqrt{3 m R_0 \omega }} \, \, .
\label{fourier of plunge J}
\eea
The transition rate out of the vacuum state is given by Fermi's golden rule \cite{hep-th/9702015,hep-th/9706100}
\bea
\mathcal{R} = \int d \omega \sum_{\ell, m} \left| J_{\ell m \omega} \right|^2 \int d \phi \, dt \, e^{i(\omega t - m \phi)} G(\phi,t) \, \, ,
\label{transition rate}
\eea
where $G(\phi,t) = \langle \mathcal{O}^{\dag}(\phi,t) \mathcal{O}(0,0) \rangle_{T_L}$ is the two point function of the CFT with left temperature $T_L=1/(2\pi)$ and an angular potential \cite{Bredberg:2009pv}.  Performing the integrals over $t,\phi$ with the appropriate $i \epsilon$  prescription and the operator normalization $C_{\mathcal{O}} = 2^{h-1} (2h-1) M/(2\pi)$ found in $\cite{Porfyriadis:2014fja}$, we find perfect agreement with the bulk flux computation:
\bea
\mathcal{R}_{\ell m \omega} =  \frac{M^2}{8 \pi} \left|\frac{X}{W_{\Omega}}\right|^2 \sqrt{\frac{3 m R_0}{\omega }} e^{\pi  m} e^{-2 \sqrt{3 m R_0 \omega }}  = \mathcal{F}_{\ell m \omega} \, \, .
\label{CFT rate}
\eea

\subsection{Gluing to asymptotically flat region}\label{matching to far region}

In section \ref{solution in NHEK} we computed a particular solution for the scalar field with Neumann (reflecting) boundary conditions at the boundary of NHEK. Here we reattach the asymptotically flat region of extreme Kerr while allowing radiation to leak outside of NHEK, imposing a boundary condition of purely outgoing waves at future null infinity in the full Kerr spacetime (keeping $\Psi = 0$ at the past horizon). We give only partial details of the matched asymptotic expansions procedure; for more details, see \cite{Porfyriadis:2014fja}.

We expand the scalar field on extreme Kerr in Boyer-Lindquist coordinate (\ref{Kerr}) modes:
\bea
\Psi = \frac{1}{\sqrt{2\pi}} \int d \hat{\omega} \sum_{\ell, m} e^{i(m\hat{\phi}-\hat{\omega} \hat{t})} \, \hat{S}_{\ell}(\theta) \, \hat{R}_{\ell m \hat{\omega}}(\hat{r}) \, \, ,
\label{scalar expansion in Kerr}
\eea
where $\hat{S}_{\ell}(\theta)$ are extreme Kerr spheroidal harmonics and $\hat{R}_{\ell m \hat{\omega}}(\hat{r})$ are radial wavefunctions. Modes penetrate the near-horizon region only if their frequency is close to the superradiant bound
\bea
\omega = 2 M \hat{\omega} - m \ll 1 \, \, .
\label{superradiant condition}
\eea
In this case, to leading order in $\omega$, the extreme Kerr spheroidal harmonics identify with those of NHEK, $\hat{S}_{\ell}(\theta) = S_{\ell}(\theta)$.
The radial wavefunction admits two useful approximations. The \emph{near-horizon} region $r \ll 1$ corresponds to NHEK: the solution of section \ref{solution in NHEK} is valid there. In the \emph{far} region $r \gg \omega$ the radial equation simplifies, allowing again a solution in terms of Whittaker functions. For our choice of boundary conditions, the far region solution behaves asymptotically as
\bea
\hat{R}^{far}_{\ell m \hat\omega}(r \to \infty) &=& Q \, \frac{\Gamma(2-2h)}{\Gamma(1-h+im)} \, (im)^{h-1+im} \times \, \, \non \\
&&\quad\times \left[ 1-\frac{(-im)^{2h-1} \sin\pi(h+im)}{(im)^{2h-1} \sin\pi(h-im)} \right] r^{-1+im} \, e^{imr/2} \, \, , \non \\
\hat{R}^{far}_{\ell m \hat\omega}(r \to 0) &=& Q \, \left[ r^{-h} -(-im)^{2h-1} \, \frac{\Gamma(2-2h) \Gamma(h-im)}{\Gamma(2h) \Gamma(1-h-im)} \, r^{h-1} \right] \, \, ,
\label{far asymptotic behaviors}
\eea
where $Q$ is a coefficient determined by the matching condition.

Gluing the asymptotically flat region is done by matching of asymptotic expansions in the near and far regions. In the near region the desired solution is found by adding to the particular solution of section \ref{solution in NHEK} the ingoing homogeneous solution $R_{hom} = W_{im,h-1/2}\left({-2i\omega}/{r} \right)$, with the appropriate amplitude $\mathcal{C}$, such that the horizon boundary conditions are not spoiled but radiation is allowed to leak outside the NHEK boundary. The matching condition then
determines the constants $Q,\mathcal{C}$ and fixes the wavefunction uniquely. In particular, the waveform at infinity, for $m>0$, $\omega>0$ and real $h<1$, is given by
\bea
&&\hat{R}_{\ell m \hat{\omega}}(r \to \infty) = {\lambda\over\pi} \, \sqrt{\frac{R_0}{2\pi}} \, \, 2^{h}\, ie^{i\pi h} \, \sin(4\pi h) \, \, e^{\pi m/2} \, \,  m^{h+im-3/2} \, \, S_{\ell}(\pi/2) \,  W_{im,h-1/2}(3im/2)\times \non \\
&&\qquad\qquad\quad\times  \frac{\Gamma(1-2h) }{\Gamma(2h-1)}\Gamma(h-i m)^2 \, \omega^{h-1/2} \, K_{2h-1}\left(\sqrt{3 m R_0 \omega}\right) \times \\
&&\quad\times     \left[ 1-(-2i\omega)^{2h-1} (-im)^{2h-1}\frac{\Gamma(1-2h)^2 \Gamma(h-im)^2}{\Gamma(2h-1)^2 \Gamma(1-h-im)^2} \right]^{-1} r^{-1+im} e^{imr/2} \, \, .\non
\label{rad at infty}
\eea

\section{Radiation from the fast near-NHEK plunge}\label{Fast near-NHEK plunge}

The near-NHEK geometry may be obtained by starting with a near-extremal Kerr BH characterized by
\bea
a=M\sqrt{1-(\kappa\epsilon)^2}\,,
\label{kappa def}
\eea
making the coordinate transformation \eqref{NHEK zoom} and taking the limit $\epsilon\to 0$. This yields the nonsingular near-horizon metric near-NHEK \cite{Bredberg:2009pv}
\bea
ds^2 = 2M^2 \Gamma(\theta)\left[ -r(r+2\kappa) dt^2 + \frac{dr^2}{r(r+2\kappa)} + d\theta^2 + \Lambda(\theta)^2 (d\phi + (r+\kappa) dt)^2 \right] \, \, .
\label{near-NHEK 1}
\eea

\subsection{Trajectory \& mapping}\label{trajectory and mapping nN}
Fast near-NHEK plunges with energy and angular momentum (per unit rest mass)
\bea
e=\frac{2 M \kappa}{\sqrt{3} R_0} \, \, , ~~~~ l = \frac{2M}{\sqrt{3}} \, \, ,
\label{e and l nN}
\eea
for arbitrary $R_0>0$, are given by
\bea \label{trajectory solution nN}
t(r) &=& \frac{1}{2\kappa} \ln\frac{r+\kappa(1+R_0) + \sqrt{\kappa} \sqrt{2 R_0 r + \kappa (1+R_0)^2}}{r+\kappa(1+R_0) - \sqrt{\kappa} \sqrt{2 R_0 r + \kappa (1+R_0)^2}} \, \, , \non \\
\phi(r) &=& \frac{3}{4\sqrt{\kappa}}\sqrt{2 R_0 r + \kappa (1+R_0)^2} \non \\
&+& \frac{1}{2}\ln\frac{R_0 r+\kappa(1+R_0) - \sqrt{\kappa} \sqrt{2 R_0 r + \kappa (1+R_0)^2}}{R_0 r+\kappa(1+R_0) + \sqrt{\kappa} \sqrt{2 R_0 r + \kappa (1+R_0)^2}} + \Phi_0 \, \, .
\eea
\begin{figure}[!hb]
\includegraphics[angle=0, width=1.6in]{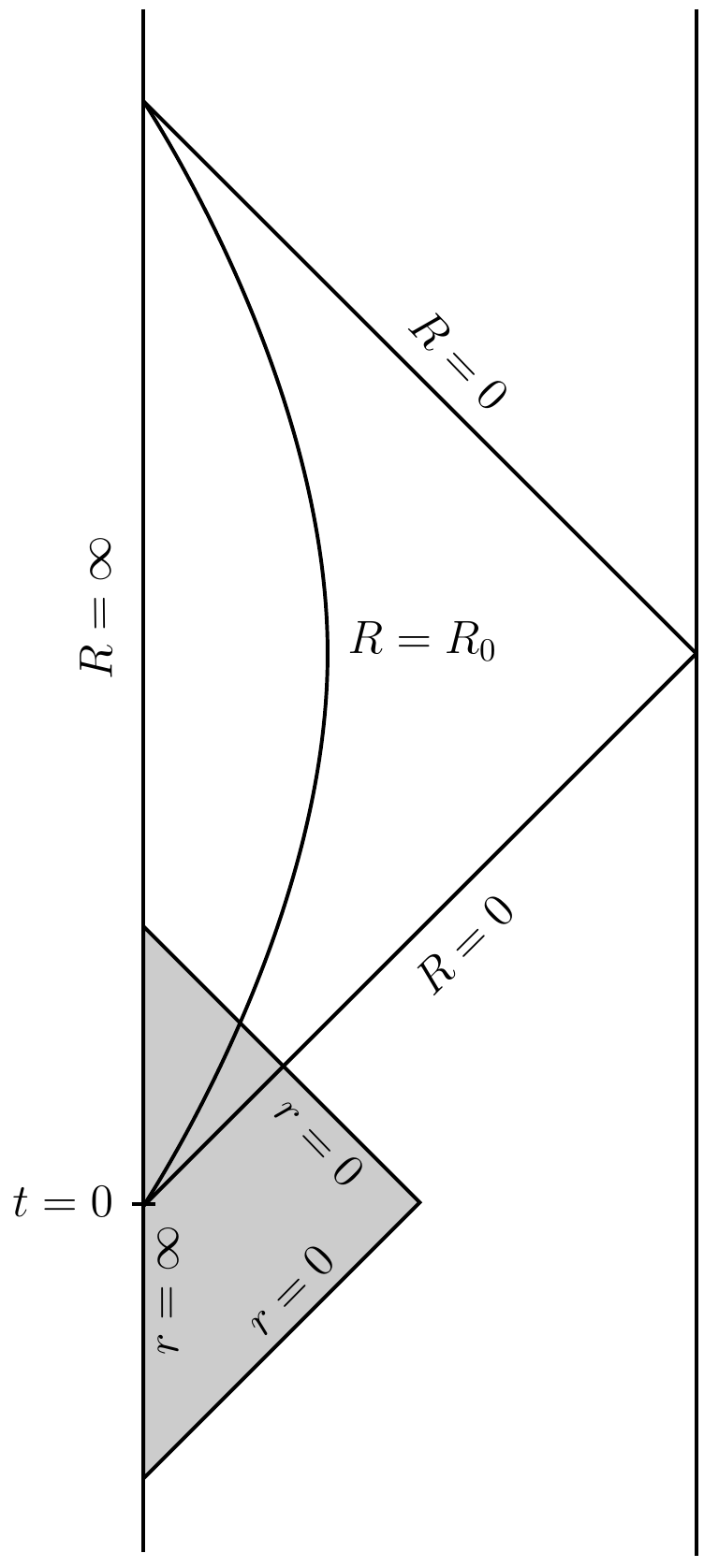}\centering
\caption{Penrose diagram with the fast near-NHEK plunge. The upper (bounded by $R=0$ and $R=\infty$) wedge is NHEK. The lower (bounded by $r=0$ and $r=\infty$) shaded wedge is near-NHEK. The line $R=R_0$ is a circular orbit in the upper wedge. In the shaded wedge it plunges from the boundary $r=\infty$ at $t=0$ into the near-NHEK future horizon $r=0$.}\label{near-NHEK diagram}
\end{figure}
The transformation
\bea
T &=& - \frac{\sqrt{r(r+2\kappa)} \sinh \kappa t}{\sqrt{r(r+2\kappa)} \cosh\kappa t-(r+\kappa)} \, \, , \non \\
R &=& \frac{1}{\kappa} \left[ \sqrt{r(r+2\kappa)} \cosh\kappa t - (r+\kappa) \right] \, \, ,  \label{mapping fast near-NHEK}  \\
\Phi &=& \phi - \frac{1}{2}\ln\frac{\sqrt{r(r+2\kappa)} - (r+\kappa) \cosh\kappa t + \kappa \sinh\kappa t}{\sqrt{r(r+2\kappa)} - (r+\kappa) \cosh\kappa t - \kappa \sinh\kappa t} \non \, \, ,
\eea
maps the $\kappa t+\ln\sqrt{r/(r+2\kappa)} \geq 0$ part of the near-NHEK geometry (\ref{near-NHEK 1}) to the $T + {1}/{R} \leq -1$ part of the geometry (\ref{NHEK 2}).
The fast plunge trajectory (\ref{trajectory solution nN}) is mapped under \eqref{mapping fast near-NHEK} to the circular NHEK orbit (\ref{circular orbit}) (see Figure \ref{near-NHEK diagram}).

\subsection{Gravity analysis}\label{solution in near-NHEK}

Consider a fast near-NHEK plunging scalar-charged star with trajectory (\ref{trajectory solution nN}). This system is described by the action (\ref{scalar action}) with source term (\ref{scalar source}).
The solution in the case of a circular orbit in NHEK (\ref{circular orbit}), as found in \cite{Porfyriadis:2014fja}, was given in section \ref{solution in NHEK}. Here, we compute the radiation produced by the fast near-NHEK plunge (\ref{trajectory solution nN}), using the mapping \eqref{mapping fast near-NHEK}. Performing this transformation on the solution (\ref{circular orbit solution}) gives the desired solution here (setting $\Phi_0=0$). As in the fast NHEK plunge, the condition of no-incoming radiation from the past horizon in (\ref{near-NHEK 1}) implies no-incoming radiation from the past horizon in (\ref{NHEK 2}) (see Figure \ref{near-NHEK diagram}).

\noindent
\textbf{Solution at the boundary.}
For fixed $t,\phi$ and $r \to \infty$
\bea
R &\approx& \frac{2r}{\kappa} \, \sinh^2{\kappa t\over 2} \to \infty \, \, , \non \\
T &\approx& -\coth\frac{\kappa t}{2} \, \, , \non \\
\Phi &\approx& \phi \, \, .
\label{mapping infinity nN}
\eea
Putting this together and using (\ref{whittaker M infty}), the solution near the boundary, for $t>0$, is
\bea
\left. \Psi \right |_{r \to \infty} = \left[ \left({\kappa\over 2}\right)^h \frac{Z}{W_{\Omega}} (-2i\Omega)^h \right] \frac{1}{\sqrt{2\pi}} \int d \omega \sum_{\ell, m} N_{\ell m \omega} \, e^{i (m \phi - \omega t) } S_{\ell}(\theta) \, r^{-h}  \, \, ,
\label{particular sol at boundary freq nN}
\eea
with
\bea
N_{\ell m \omega} \, =  \, \frac{1}{\sqrt{2\pi}} \int_{0}^{\infty} dt \, e^{i \omega t} \, e^{- \frac{3 i m R_0}{4} \coth\frac{\kappa t}{2}}\left(\sinh{\kappa t\over 2}\right)^{-2h}   \, \, .
\label{fourier infty nN}
\eea
\noindent
\textbf{Solution at future horizon.}
For $r \to 0$ with fixed $v \equiv \kappa t + \ln\sqrt{{r}/{(r+2\kappa)}}$ and $\phi$,
\bea
R &\approx& e^v - 1 \, \, , \non \\
T &\approx& -\frac{1}{1-e^{-v}} \, \, , \non \\
\Phi &\approx& \phi - \ln(1-e^{-v}) - \ln\sqrt{\frac{r}{2 \kappa}} \, \, .
\label{mapping horizon nN}
\eea
The solution near the horizon, for $v>0$, is given by
\bea
\Psi_{hor} = \sum_{\ell, m} e^{im \phi} S_{\ell}(\theta) \left( \frac{r}{2 \kappa} \right)^{-\frac{i m}{2}}  \left( 1-e^{-v} \right)^{ - i m} e^{-\frac{3 i m R_0}{4(1 - e^{-v})}} R^{(c)}_{\ell m}(e^v - 1) \, \, .
\label{particular sol at horizon nN}
\eea
In terms of the Fourier decomposition
\bea
\Psi_{hor} &=& \frac{1}{\sqrt{2\pi}} \int d\omega \sum_{\ell, m} e^{i(m \phi - \omega t)} S_{\ell}(\theta) \, R^{hor}_{\ell m \omega}(r) \, , \label{fourier horizon1 nN} \non \\
R^{hor}_{\ell m \omega} &=&  \left( \frac{r}{2 \kappa} \right)^{-\frac{i}{2} \left(\frac{\omega }{\kappa }+m\right)} \frac{X}{W_{\Omega}} \frac{\kappa^{-1}}{\sqrt{2 \pi}} \int_{0}^{\infty} dv \, e^{i {\omega\over\kappa} v} \left( 1-e^{-v} \right)^{ - i m} e^{-\frac{3 i m R_0}{4(1 - e^{-v})}} W_{im,h-1/2}\left( \frac{3 i m R_0}{2 (e^v - 1)} \right)  \, \, , \non
\label{fourier horizon2 nN}
\eea
where as in (\ref{fourier horizon2}) we consider the high frequency limit $\omega/\kappa \gg 1$ and disregard the $v>\ln(1+R_0)$ part of the Fourier integral. The remaining integral above may be evaluated approximately using $\omega / \kappa \gg 1$ to obtain, for $\omega,m>0$ and real $h<1$,
\bea
R^{hor}_{\ell m \omega} &\approx& \frac{i}{2} \, \frac{X}{W_\Omega}  \, 2^{\frac{i}{2} \left(\frac{\omega }{\kappa }+m\right)} \, \left( \frac{6 m R_0 \omega}{\kappa} \right)^{1/4} \, \omega^{-1 + i m} \, \kappa^{\frac{i}{2}\left(\frac{\omega }{\kappa} -m\right) } \,  e^{\frac{ \pi m}{2}} \, e^{-\sqrt{\frac{6 m R_0 \omega}{\kappa}}}\, r^{-\frac{i}{2} \left(\frac{\omega }{\kappa }+m\right)}  \, \, .\non
\label{fourier horizon3 nN}
\eea
The particle number flux down the horizon is thus given, for $\omega / \kappa \gg 1$, by
\bea
\mathcal{F}_{\ell m \omega} \approx \frac{M^2}{4 \pi} \left| \frac{X}{W_\Omega} \right|^2 \sqrt{\frac{3 m R_0}{2 \omega \kappa}} \, e^{\pi m} \, e^{-2\sqrt{\frac{6 m R_0 \omega}{\kappa}}} \, \, .
\label{flux nN}
\eea

\subsection{CFT analysis}\label{CFT analysis nN}
The transformation \eqref{mapping fast near-NHEK} induces the following conformal transformation on the boundary
\bea
T &=& - \coth \frac{\kappa t}{2} \, \, ,  \non \\
\Phi &=& \phi \, \, .
\label{boundary conformal transformation nN}
\eea
Note that this is not one of the global $SL(2,R)$ transformations, instead it is one of the local Virasoro symmetries of Kerr/CFT. After this conformal transformation the CFT source \eqref{circular J} becomes
\bea
J_{\ell}(\phi,t) = \Theta(t) \left( \frac{\kappa}{2} \right)^{1-h} \left( \sinh\frac{\kappa t}{2} \right)^{2h-2} \sum_{m} \frac{X}{W_{\Omega}} (-2i \Omega)^{1-h} \frac{\Gamma(2h-1)}{\Gamma(h-im)} \, e^{i m \phi + i \Omega \coth \frac{\kappa t}{2}} \, \, .
\label{plunge J nN}
\eea
The Fourier decomposition in the $\omega / \kappa \gg 1$ limit, for $\omega,m>0$ and real $h<1$, is
\bea
J_{\ell}(\phi,t) &=& \frac{1}{\sqrt{2 \pi}} \int d\omega \sum_{m} J_{\ell m \omega} e^{i(m\phi - \omega t)} \, \, , \non \\
J_{\ell m \omega} &\approx& \frac{X}{W_{\Omega}} \, \frac{\Gamma (2 h-1)}{\Gamma (h-i m)} \, \left(\frac{6 m R_0 \omega }{\kappa }\right)^{1/4} \, (2\omega)^{-h} \, e^{\frac{i \pi  h}{2} - \sqrt{\frac{6 m R_0 \omega }{\kappa }}} \, \, .
\label{fourier of plunge J nN}
\eea
The transition rate out of the vacuum state, given by Fermi's golden rule \eqref{transition rate} with
$G(\phi,t) = \langle \mathcal{O}^{\dag}(\phi,t) \mathcal{O}(0,0) \rangle_{T_L,T_R}$ the two point function of the CFT at $T_L = \frac{1}{2\pi}$, $T_R = \frac{\kappa}{2\pi}$ and an angular potential. Performing the integrals, we find for $\omega/\kappa\gg 1$,
\bea
\mathcal{R}_{\ell m \omega} =  \frac{M^2}{4 \pi} \left|\frac{X}{W_{\Omega}}\right|^2 \sqrt{\frac{3 m R_0}{2  \omega  \kappa}} e^{\pi  m} e^{-2 \sqrt{\frac{ 6 m R_0 \omega}{\kappa} }}  = \mathcal{F}_{\ell m \omega} \, \, .
\label{CFT rate nN}
\eea
which establishes perfect agreement with the bulk gravity computation.

\subsection{Gluing to asymptotically flat region}\label{matching to far region nN}

To compute radiation at future null infinity, we reattach the asymptotically flat region of the full near-extremal Kerr to our near-NHEK and glue solutions via matched asymptotic expansions. This is done in a similar manner to that described in section \ref{matching to far region} for exactly extremal Kerr. Details of the procedure may be found in \cite{Hadar:2014dpa}; here we will directly state the final result.

We expand the scalar field in Kerr as in (\ref{scalar expansion in Kerr}).
Via the dimensionless Hawking temperature $\tau_H = (r_+ - r_-)/r_+$ and horizon angular velocity $\Omega_H = a/(2M r_+)$, we define a rescaled near-superradiant frequency
\bea
n \equiv 4 M \, \frac{\hat{\omega}-m \Omega_H}{\tau_H} \, \, .
\label{near extremal kerr rescaled frequency}
\eea
We can then perform matched asymptotic expansions for $\textrm{max}(\tau_H, n\tau_H)\ll 1$. The waveform at future null infinity, for $m>0$, is given by
\bea
&&\hat{R}_{\ell m \hat{\omega}}(r \to \infty) =  {\lambda \over \pi}  R_0^h  \,  2^{2-3h} \, 3^{h-1/2} \, \tau_H^{h-1} \, \sin(2\pi h) \, e^{\pi m /2} \, m^{2h+im-2}  \times \non \\
&&\qquad\qquad\quad  \times \, S_{\ell}(\pi/2) \,  W_{im,h-1/2}(3im/2) \, \,  \frac{ \Gamma(1-2h) }{\Gamma(2h-1)}\Gamma(h-i m)^2 \, \mathcal{N} \times  \\
&&\quad \times  \left[ 1-(-i m \tau_H)^{2h-1} \frac{\Gamma(1-2h)^2 \Gamma(h-im)^2}{\Gamma(2h-1)^2 \Gamma(1-h-im)^2} \frac{\Gamma(h-i(n-m))}{\Gamma(1-h-i(n-m))} \right]^{-1} r^{-1+im} e^{imr/2} \, \, , \non
\label{rad at infty nN}
\eea
where $\mathcal{N}$ is the integral (\ref{fourier infty nN}):
\bea
\mathcal{N}=\frac{1}{\sqrt{2\pi}} \int_{0}^{\infty} dy \, e^{i (n-m) y} \, e^{- \frac{3 i m R_0}{4} \coth\frac{y}{2}}\left(\sinh{y\over 2}\right)^{-2h}\,\,.
\eea

\section*{Acknowledgements}
APP and AS were supported in part by DOE grant DE-FG02-91ER40654. SH was supported by the Israel Science Foundation, grant no. 812/11, and by the Einstein Research Project ``Gravitation and High Energy Physics'', which is funded by the Einstein Foundation Berlin.

\end{document}